\documentclass[preprint,12pt]{elsarticle}

\usepackage{amssymb}

\usepackage{bbm}
\usepackage{xcolor}
\usepackage{amsmath}
\usepackage{url}
\usepackage{hyperref}
\usepackage{graphicx}

\newtheorem{lemma}{Lemma}[section]
\newtheorem{proposition}{Proposition}[section]
\newtheorem{theorem}{Theorem}[section]
\newtheorem{definition}{Definition}[section]

\newenvironment{profTheorem}[1][Proof of Theorem ]{\textbf{#1} }{\ \rule{0.5em}{0.5em}}
\newenvironment{profProposition}[1][Proof of Proposition ]{\textbf{#1} }{\ \rule{0.5em}{0.5em}}
\newenvironment{profLemma}[1][Proof of Lemma ]{\textbf{#1} }{\ \rule{0.5em}{0.5em}}
\newproof{pf}{Proof}
\newproof{pot}{Proof of Theorem \ref{thm2}}

\journal{Computational Statistics \& Data Analysis}

\begin{document}

\begin{frontmatter}


\title{Title\tnoteref{label1}}

\title{Model-Based Clustering using multi-allelic loci data with loci selection}


\author[addr1,addr2,addr3]{Wilson TOUSSILE\corref{cor1}}
\ead{Wilson.Toussile@math.u-psud.fr}
\ead[url]{http://www.math.u-psud.fr/\~{}toussile}
\cortext[cor1]{Wilson Toussile, Laboratoire de Math\'ematiques d'Orsay, B\^at 425, 91405 Orsay cedex, Tel : +33 (0) 1 69 15 76 18, +33 (0) 6 12 98 44 43}

\address[addr1]{UR016, Institut de Recherche pour le D\'eveloppement (IRD)}
\address[addr2]{Ecole Nationale Sup\'erieure Polytechnique (ENSP), Universit\'e de Yaound\'e 1, Cameroun}
\address[addr3]{Laboratoire de Math\'ematiques d'Orsay (LMO), UMR 8628, Universit\'e Paris-Sud 11}

\author[addr3]{Elisabeth GASSIAT}
\ead{Elisabeth.Gassiat@math.u-psud.fr}
\ead[url]{http://www.math.u-psud.fr/\~{}gassiat}

\begin{abstract}
  A long standing issue in population genetics is the identification of genetically homogeneous populations. The most widely used measures of population structure are Wright's F statistics (Wright 1931). But the fundamental prerequisite of any inference based on these statistics is the definition of populations and this definition is typically subjective (based on linguistic, cultural or physical characters, geographical location). The population structure may be difficult to detect using visible characters.

  We propose a Model-Based Clustering (MBC) method combined with loci selection using multi-allelic loci data. The loci selection problem is regarded as a model selection problem and models in competition are compared with the Bayesian Information Criterion (BIC). The resulting procedure selects the subset $\widehat{S}_n$ of clustering variables, the number $\widehat{K}_n$ of clusters, estimates the proportion of each population and the allelic frequencies within each cluster. We prove that the selected model $\left(\widehat{K}_n, \widehat{S}_n\right)$ converges in probability to the true model $\left(K_0, S_0\right)$ under a single realistic assumption as the size $n$ of the sample tends to infinity. The proposed algorithm named \textbf{MixMoGenD} ('Mixture Model for Genetic Data') has been implemented using $C++$ and $C$ programming languages. An interface with \textbf{R} was created. Numerical experiments on simulated data sets was conducted to highlight the interest of the proposed loci selection procedure.
\end{abstract}

\begin{keyword}
  Model-Based Clustering \sep Model Selection \sep Variable Selection \sep BIC \sep Population Genetics



\end{keyword}

\end{frontmatter}



\section{Introduction}
\label{introduction}

  Evolutionary processes have produced an immense array of biological diversity on our planet, with species displaying complex adaptations to their environments. Understanding this diversity and complexity, its origins, and its implications is a daunting challenge. Population genetics provides tools to meet this challenge. It is concerned with origin, amount, and distribution of genetic variation present in populations of organisms and fate this variation through space and time. A long standing issue in population genetics is the identification of genetically homogeneous populations. The most widely used measures of population structure are Wright's F statistics (Wright 1931). But the fundamental prerequisite of any inference based on these statistics is the definition of populations and this definition is typically subjective, based on linguistic, cultural or physical characters, or geographic location. The population structure may be difficult to detect using visible characters.
  \vspace{0.10cm}

  This article is concerned with population structure that is difficult to detect using visible characters (such as linguistic, cultural, physical characters, or geographic location), but may be significant in genetic terms. For example, the problem of cryptic population arises in the context of DNA fingerprinting for forensics, where it is important to assess the degree of population structure to estimate the probability of false matches (\emph{D. J. Balding and Nichols RA} 1994 et 1995 \cite{Balding1994}, \cite{Balding1995a}, \cite{Balding1995}; \emph{Foreman et al.} (1997) \cite{Foreman1997}). 
  \vspace{0.10cm}
  
  Several Model Based-Clustering (MBC) for multi-locus genetic data have been developped in recent years: \textbf{STRUCTURE} by \emph{J. K. Pritchard et al.} (2000) \cite{Pritchard2000}, \textbf{BAPS} by  \emph{J. Corander. et al.} (2004) \cite{Cor2004}, \textbf{FASTRUCT} by \emph{Olivier François et al.} (2006) \cite{Francois2006}. These methods attempt to group samples into clusters of random mating individuals so that the \emph{Hardy-Weinberg} (HW) and linkage disequilibria (LD) are minimized across the sample. Although Hardy-Weinberg and linkage equilibria models are based on several simplifying assumptions that can be unrealistic, they have still proven to be useful in describing many population genetics attributes and will serve as a useful base model in the development of more realistic models of microevolution. \textbf{STRUCTURE} and \textbf{BAPS} are bayesian methods that use MCMC algorithms and thus require much longer computations than frequentist likelihood methods using Expectation-Maximization (EM) algorithm \cite{Dempster1977}. 
  \vspace{0.10cm}
  
  Multi-locus data sets are becoming increasingly large due to the explosion of genomic projects. But, the structure of interest may be contained in only a subset of available loci, the others being useless or even harmful to detect a reasonable clustering structure. It then becomes necessary to select the optimum subset $S_0$ of loci which cluster in the best way the population. None of the methods cited above include a loci selection procedure. Defining the optimum set $S_0$ of loci and optimum number $K_0$ of subpopulations requires a suitable variable selection procedure.
  \vspace{0.10cm}
  
  In this article, we propose a new clustering method and an associated algorithm named Mixture Model for Genetic Data (\textbf{MixMoGenD}) that has three benefits. First, it is model-based. Second, it is based on EM algorithm, so it is relatively fast compared to its counterparts based on MCMC \cite{Francois2006}. The main benefit of our proposed method is that it is coupled with a loci selection procedure based on Bayesian Information Criterion (BIC) and on a backward stepwise method. Recall that in clustering, classification is not observed and there is no a priori knowledge of the structure being looked for in the analysis, and of the subset of available loci that are relevant for discrimination. So there is no simple pre-analysis screening technic available to use. Thus it makes sense to include loci selection procedure as a part of the clustering algorithm as recommended by \emph{C. Maugis et al.} (2007) \cite{Maugis2007} in a gaussian framework. The resulting procedure selects the subset $S$ of clustering variables, the number $K$ of clusters, estimates proportion and allelic frequencies within each cluster.
  \vspace{0.10cm}

  We recast loci selection problem and the estimation of the number of clusters as a model selection problem. Bayes factors, the ratio of integrated likelihood for models, are used to compare models, so that the models to be compared can be non-nested. Since  integrated likelihood is usually difficult to compute, the Bayesian Information Criterion (BIC) for the competing models is used to approximate the log-likelihood. \emph{Raftery et al.} (2006) \cite{Raftery2006} showed that compared to clustering methods based on all variables, variable selection method based on the BIC consistently yielded more accurate estimates of the number of clusters in a gaussian context. The proposed method can be applied to various types of markers (e.g. microsatellites, restriction fragment length polymorphisms (RFLPs), or single nucleotide polymorphisms (SNPs)).
  \vspace{0.25cm}

  The model and methods are presented in section \ref{mat}. The consistency of the estimators of the number of populations and the set of loci relevant for discrimination is proved in Section \ref{consistency} under a single realistic assumption. The proposed algorithm has been implemented using $C++$ and $C$ programming languages. In Section \ref{simulation}, Numerical experiments on simulated data sets was conducted to highlight the interest of the proposed loci selection procedure.
  
  The program, sample project files and their simulation parameters, and documentation for linux OS are available free of charge at :\\ \url{http://www.math.u-psud/~toussile}.
\section{Model and methods} 
\label{mat}

\vspace{0.25cm}
    In this section, we present the model and our clustering method using loci selection based on the Bayesian Information Criterion. The loci selection procedure is presented in sub-section \ref{procedure_selection}, the identifiability of models in competition and of parameters are discussed in sub-section \ref{identifiabilite}, and the EM equations are given in sub-section \ref{equations_EM}.

  \subsection{Notations and estimation method}  
  \label{Principe}
  \vspace{0.25cm}
  
    The data set we shall deal with consists of genotypes of $n$ diploid individuals labeled $1,\ldots,\ i,\ \ldots,\ n$ at $L$ loci labeled $1,\ldots,\ l,\ldots,\ L$. The observations are written as $x=\left(x_{i}^l\right)_{i=1,\ldots,n;\ l=1,\ldots,L}$, where $x_{i}^l=\left\{x_{i,1}^l,\ x_{i,2}^l\right\}$ is the genotype of the $i^{th}$ individual at the $l^{th}$ locus. The data set $x$ is assumed to be a realization of a random vector $X=\left(X_{i}^l\right)_{i=1,\ldots,n,\ l=1,\ldots,L}$, where $X_{i}^l=\left\{X_{i,1}^l,\ X_{i,2}^l\right\}$, with $X_{i,1}^l$ and $X_{i,2}^l$ taking values in the set $\left\{1,\ldots,\ l,\ldots,\ A_l\right\}$, and $1,\ldots,\ j,\ldots,\ A_l$ denote the labels of distinct alleles that are observed at locus $l$. We assume that the variables $X_i=\left(X_i^l\right)_{l=1,\ldots,\ L}$, $i=1,\ldots,\ n$ are independent and identically distributed.

    Let : 
    \begin{itemize}
      \item $z_i$ be the (unobserved) population of origin of individual $i$;
      \item $\pi_k:=P\left(z_i=k\right)$ be the proportion of population $k$;
      \item $\alpha_{k,l,j}:=P\left(X_{i,1}^l=j|\ z_i=k\right)=P\left(X_{i,2}^l=j|\ z_i=k\right)$ be the frequency of the $j^{th}$ allele at locus $l$ in population $k$;
      \item $\mathcal{X}$ be the set of possible genotypes from observed alleles;
    \end{itemize}
    and let $z=\left(z_1,\ldots,\ z_n\right)$, $\pi=\left(\pi_1,\ldots,\ \pi_K\right)$ and $\alpha=\left(\alpha_{k,l,j}\right)_{k=1,\ldots,K;\ l=1,\ldots,L;\ j=1,\ldots,A_l}$. The $\pi_k$'s are called the mixing proportions and represent the prior probability of an individual coming from each population $k$.
    \vspace{0.50cm}
    
    Our main modeling assumptions are
    \begin{enumerate}[$\left(\mathcal{H}_1\right)$:]
      \item Hardy-Weinberg Equilibrium (HWE) within populations and
      \item complete Linkage Equilibrium (LE) within populations.
    \end{enumerate}

    Model-based methods proceed by assuming that observations from each cluster are drawn from some parametric model and the overall population is a finite mixture of these populations. Thus, without loci selection, observations $x=\left(x_1,\ldots,\ x_n\right)$ are supposed to be a sample from the probability distribution with the likelihood contribution of individual $i$ given by the following equation
    \begin{equation}\label{Melange}
      P_K\left(x_i|\ \theta\right):=P\left(x_i|\ K,\ \theta\right)=\sum_{k=1}^K\pi_k\left[\prod_{l=1}^LP\left(x_i^l|\ z_i=k,\ \alpha_{k,l,\cdot}\right)\right],
    \end{equation}
    where $\theta=\left(\pi,\ \alpha\right)$ is a parameter ranging in a certain space $\Theta_K$, for a given number $K$ of populations.
    \vspace{0.50cm}
    
     In this model of probability distributions, all the $L$ loci are supposed to be relevant for clustering. Now, the structure of interest may be contained in only a subset $S$ of available loci, the others being useless or even harmful to detect a reasonable clustering structure. Let $S^c$ be the subset of loci that are irrelevant for clustering ($S\cup S^c=\left\{1,\ldots,\ L\right\}$). The natural third hypothesis is the following.
    \begin{enumerate}[$\left(\mathcal{H}_3\right)$:]
     \item the alleles of the loci of $S^c$ are identically distributed in the overall population, i.e 
      \begin{equation}
        \alpha_{1,l,j}=\alpha_{2,l,j}=\ldots=\alpha_{K,l,j}=:\beta_{l,j},\ \forall l\in S^c\ \textrm{and }\forall j\in\left\{1,\ldots,\ A_l\right\} .
      \end{equation}
    \end{enumerate}

   The allele frequencies are given by the Hardy-Weinberg model:
    \begin{equation}  \label{HW_modele}
      P\left(x_i^l|\ z_i=k,\ \alpha_{k,l,\cdot}\right) =\left(2-\mathbbm{1}_{\left[x_{i,1}^l=x_{i,2}^l\right]}\right)\alpha_{k,l,x_{i,1}^l}\times\alpha_{k,l,x_{i,2}^l}.
    \end{equation}
    Although this model makes several simplifying assumptions that are unrealistic in some cases, it has still proven to be useful in describing many population genetics attributes and will serve as a first tool in the development of more realistic models of microevolution.
    
    Under the three assumptions $\left(\mathcal{H}_1\right)$, $\left(\mathcal{H}_2\right)$ and $\left(\mathcal{H}_3\right)$, and given the number $K$ of populations and the subset $S$ of relevant loci, the observations are supposed to be realizations of a sample from a probability distribution of the form
    \begin{eqnarray} \label{Melange_selection}
     P_{\left(K,\ S\right)}\left(x_i|\ \theta\right)&:=&P\left(x_i|\ K,\ S,\ \theta\right)\nonumber\\
      &=&\left[\sum_{k=1}^K\pi_k\prod_{l\in S}P\left(x_i^l|\ z_i=k,\ \alpha_{k,l,\ \cdotp}\right)\right]\times\prod_{l\in S^c}P\left(x_i^l|\ \beta_{l,\ \cdotp}\right),
    \end{eqnarray}
    where $\theta:=\left(\pi,\ \left(\alpha_{\cdot,l,\ \cdot}\right)_{l\in S},\ \left(\beta_{l,\ \cdot}\right)_{l\in S^c}\right)$ is a multidimensional parameter ranging over some space $\Theta_{\left(K,\ S\right)}$. Each individual is assumed to originate in one of the $K$ (unknown) populations, each with its own allele frequencies. Thus these parameters verify the following properties :
    \begin{equation}  \label{condition_pi}
      \left\{ \begin{array}{l}
          0<\pi_k\leq1,\ k=1,\ldots,\ K;\\
          \sum_{k=1}^K\pi_k=1.
         \end{array}
      \right. 
    \end{equation}
    \begin{equation}  \label{condition_alpha}
      \left\{
        \begin{array}{l}
          0\leq \alpha_{k,\ l,\ a} \leq 1,\ k=1,\ldots,\ K,\ l\in S,\ a=1,\ldots,\ A_l;\\ 
          \sum_{a=1}^{A_l}\alpha_{k,\ l,\ a}=1,\ k=1,\ldots,\ K,\ l =1,\ldots,\ L.
        \end{array}
      \right.
    \end{equation}
    \begin{equation}  \label{condition_beta}
      \left\{
        \begin{array}{l}
          0\leq \beta_{l,\ a} \leq 1,\ l\in S^c,\ a=1,\ldots,\ A_l;\\ 
          \sum_{a=1}^{A_l}\beta_{l,\ a}=1,\ l\in S^c.
        \end{array}
      \right.
    \end{equation}
    \vspace{0.50cm}

    The number $K$ of populations, the subset $S$ of relevant loci for clustering, the proportions $\pi$ of populations, the allele frequencies $\alpha$ and $\beta:=\left(\beta_{l,j}\right)_{l\in S^c;\ j=1,\ldots,\ A_l}$ are treated as the parameters of the model, which have to be infered. The variable $z_i$, the assignment of individual $i$ to its population is not observed and has to be predicted.
    \vspace{0.50cm}
    
    Infering on $K$ and $S$ is regarded as a model selection problem. In fact, each value of $\left(K,\ S\right)$ defines a parametric model  \begin{equation*}
      \mathcal{M}_{\left(K,\ S\right)}=\left\{P_{\left(K,\ S\right)}\left(\ \cdot\ |\ \theta\right);\ \theta\in\Theta_{\left(K,\ S\right)}\right\}
    \end{equation*}
    of probability distributions. Let $K_{\max}$ be the maximum number of clusters. $K_{\max}$ has to be specified by the user for identifiability purposes discussed in the sub-section \ref{identifiabilite} hereafter. Let us consider the collection $\mathcal{C}$ of competing models:
    \begin{equation}  \label{collection_modeles}
      \mathcal{C}=\left\{\mathcal{M}_{\left(K,\ S\right)}\ :\ K\in\left\{1,\ldots,\ K_{\max}\right\}\ \textrm{and }S\in\mathcal{P}^{*}\left(L\right)\right\},
    \end{equation}
    where $\mathcal{P}^{*}\left(L\right)$ is the set of non-empty subsets of the available loci $\left\{1,\ldots,\ L\right\}$.
    
    In a Bayesian framework, the model $\mathcal{M}_{\left(\widetilde{K}_n,\ \widetilde{S}_n\right)}$ maximizing the posterior probability is to be chosen :
    \begin{equation}
      \left(\widetilde{K}_n,\ \widetilde{S}_n\right) = \arg\max_{\left(K,\ S\right)}P\left[\left(K,\ S\right)|\ x\right].
    \end{equation}
    By Bayes Theorem and assuming a non informative uniform prior distribution $P\left[\left(K,\ S\right)\right]$ on the competing models $\mathcal{M}_{\left(K,\ S\right)},\ K=1,\ldots,\ K_{\max},\ S\in\mathcal{P}^{*}\left(L\right)$, one  has 
    \begin{equation}
      \left(\widetilde{K}_n,\ \widetilde{S}_n\right)=\arg\max_{\left(K,\ S\right)}P\left[x|\ \left(K,\ S\right)\right].
    \end{equation}
    The quantity  $P\left[x|\ \left(K,\ S\right)\right]$ is the integrated likelihood of model $\mathcal{M}_{\left(K,\ S\right)}$, namely 
    \begin{equation}  \label{vrais_integre}
      P\left[x|\ \left(K,\ S\right)\right]=\int_{\theta\in\Theta_{\left(K,\ S\right)}}\bigg(\prod_{i=1}^{n} P_{\left(K,\ S\right)}\left(x_{i}|\ \theta\right)\bigg)P\left[\theta|\ \left(K,\ S\right)\right]d\theta,
    \end{equation}
    where $d\theta$ is a measure on the parameter space $\Theta_{\left(K,\ S\right)}$ and $P\left[\theta|\ \left(K,\ S\right)\right]$, the prior distribution (\emph{Kass and Raftery 1995} \cite{Kass1995}). This integrated likelihood is analytically difficult to compute.  An asymptotic approximation of $2\ln P\left[x|\ \left(K,\ S\right)\right]$ is generally used; this approximation is the Bayesian Information Criterion (BIC) defined by 
    \begin{equation}  \label{BIC}
      BIC\left(K,\ S\right)=2\sum_{i=1}^{n}\ln P_{\left(K,\ S\right)}\left(x_{i}|\ \widehat{\theta}_{ML,(K,S)}\right)-d_{\left(K,\ S\right)}\ln n,
    \end{equation}
    where $d_{\left(K,\ S\right)}$ is the dimension of the parameter space $\Theta_{\left(K,\ S\right)}$ and $\widehat{\theta}_{ML,(K,S)}$, the maximum likelihood estimate of $\theta$ in $\Theta_{\left(K,\ S\right)}$. Thus, the selected model is given by 
    \begin{equation}
      \left(\widehat{K}_n,\ \widehat{S}_n\right)=\arg\max_{\left(K,\ S\right)}BIC\left(K,\ S\right).
    \end{equation}
    
    The maximum likelihood estimate $\widehat{\theta}_{ML,\left(\widehat{K}_n,\ \widehat{S}_n\right)}$ yields the Maximum a Posteriori (MAP) prediction rule defined by 
    \begin{equation}  \label{MAP}
      \widehat{z}_i=\arg\max_{k\in\left\{1,\ldots,\ \widehat{K}\right\}}\widehat{\pi}_kP\left(x_i|\ z_i=k,\ \widehat{\theta}_{ML,\left(\widehat{K}_n,\ \widehat{S}_n\right)}\right).
    \end{equation}
    One can notice that $\widehat{\theta}_{ML,\ (K,\ S)}=\left(\widehat{\gamma}_{ML,\ (K,\ S)},\ \widehat{\beta}_{ML,(K,S)}\right)$, where $\gamma=\left(\pi,\ \alpha\right)$.
    The maximum likelihood estimate $\widehat{\gamma}_{ML,(K,S)}$ is computed using the Expectation Maximization (EM) algorithm (\emph{Dempster et al.} (1977) \cite{Dempster1977}), and the likelihood estimate $\widehat{\beta}_{ML,(K,S)}$ is given by the observed frequencies of the alleles of the loci of $S^c$.
    \vspace{0.50cm}
    
    Before giving the EM equations, let us present the loci selection procedure.
    
  \subsection{Combined Loci selection and clustering procedure} 
  \label{procedure_selection}

  The space of competing models can be very large, consisting of all combinations of all $\left(2^L-1\right)$ non-empty subsets of the available loci with each possible number of populations. Thus an exhaustive research of an optimum model is very painful in most situations. We adopt a two nested-step algorithm as proposed by \emph{C. Maugis et al.} (2007) \cite{Maugis2007} :
  \begin{enumerate}[Step 1. ]
    \item For all $K\in\left\{1,\ldots,\ K_{\max}\right\}$, we reseach
      \begin{equation}
        \widehat{S}_n\left(K\right)=\arg\max_{S\in\mathcal{P}^{*}\left(L\right)}BIC\left(K,\ S\right)
      \end{equation}
      by a backward stepwise procedure detailed hereafter.
    \item We determine
      \begin{equation}
        \widehat{K}_n=\arg\max_{K\in\left\{1,\ldots,\ K_{\max}\right\}}BIC\left(K,\ \widehat{S}_n\left(K\right)\right).
      \end{equation}
  \end{enumerate}
  \vspace{0.50cm}
  
  We prefer a backward stepwise procedure rather than a forward stepwise as in \emph{Kass and Raftery} (1995) \cite{Kass1995} because starting the selection algorithm with all loci included allows the model to take loci interactions into account. At each stage, the algorithm searches for a locus to remove, and then assesses whether one of the current irrelevant loci can be selected. Thus the algorithm is making use of an exclusion and an inclusion procedures described hereafter. The decision of excluding or including a locus from the set of clustering loci is based on the BIC approximation of the Bayes factor.
  
  \paragraph{Backward Stepwise selection procedure}
  \vspace{0.10cm}
  \begin{description}
    \item[1 - Initialisation :] $S=\left\{1,\ldots,\ L\right\}$, $S^c=\emptyset$.
    \item[2 - Exclusion step :] The proposed locus for removal from the currently selected clustering loci $S$ is chosen to be the one from this set without which the model is the best among the models with $\sharp S-1$ loci.
    \begin{equation}
      c_{ex}=\arg\max_{l\in S}BIC\left(K, S\smallsetminus\left\{l\right\}\right).
    \end{equation}
    This candidate $c_{ex}$ is excluded if the model $\left(K,\ S\smallsetminus\left\{c_{ex}\right\}\right)$ is better than $\left(K,\ S\right)$, i.e. 
    \begin{equation}
      BIC\left(K,\ S\right)-BIC\left(K,\ S\smallsetminus\left\{c_{ex}\right\}\right)\leq 0.
    \end{equation}
    \item[3 - Inclusion step :] The proposed new clustering locus for inclusion in the currently selected clustering loci set $S$ is chosen to be the one from the set $S^c$ of currently non-selected loci which shows most evidence of multivariate clustering including the previous selected loci. This locus is accepted as relevant for clustering if its evidence for clustering is stronger than not clustering, namely
    \begin{equation}
      BIC\left(K,\ S\cup\left\{c_{in}\right\}\right)-BIC\left(K,\ S\right)>0,
    \end{equation}
    where
    \begin{equation}
      c_{in} = \arg\max_{l\in S^c}BIC\left(K,\ S\cup\left\{l\right\}\right).
    \end{equation}
  \end{description}

The algorithm repeats \textbf{2} and \textbf{3} and stops when the proposed candidate for inclusion is the locus removed in the previous step or when $S^c$ is empty.

  \subsection{EM equations} \label{equations_EM}

  \vspace{0.25cm}
  Here we describe the EM equations. To assign individual $i$ to a cluster, we compute the posterior assignment probabilities $\tau_{ik}=P\left(z_i=k|\ x_i\right)$. Hereafter, we write $\gamma^{\left(r\right)}=\left(\pi^{\left(r\right)},\ \alpha^{\left(r\right)}\right)$ for the estimate of $\gamma=\left(\pi,\ \alpha\right)$ at iteration $r$ of the EM algorithm. The $\tau_{ik}^{\left(r\right)}$ can be describe as 
  \begin{equation}
    \tau_{ik}^{\left(r\right)}=\frac{\pi_{k}^{\left(r\right)}\prod_{l\in S}P\left(x_i^l|\ z_i=k,\ \alpha_{k,l,\ \cdot}^{\left(r\right)}\right)}{\sum_{h=1}^{K}\pi_h^{\left(r\right)}\prod_{l\in S}P\left(x_i^l|\ z_i=k,\ \alpha_{h,l,\ \cdot}^{\left(r\right)}\right)}
  \end{equation}
  Then the update formulae for the parameters can be derived using the standard method of the EM algorithm
  \begin{equation}
    \pi_{k}^{\left(r+1\right)}=\frac{1}{n}\sum_{i=1}^{n}\tau_{ik}^{\left(r\right)}
  \end{equation}
  and
  \begin{equation}
    \alpha_{k,l,j}^{\left(r+1\right)}=\frac{\sum_{i=1}^{n}\tau_{ik}^{\left(r\right)}\left(\mathbbm{1}_{\left[x_{i,1}^l=j\right]}+\mathbbm{1}_{\left[x_{i,2}^l=j\right]}\right)}{2\sum_{i=1}^n\tau_{ik}^{\left(r\right)}}.
  \end{equation}
  
  When applying the EM algorithm to data, we need to provide values for $\tau_{ik}^{\left(0\right)}$. There are mainly two types of initialization methods : random initialization methods and clustering-based initialization methods (McLachlan \& Peel 2000). The random initialization methods assign individuals into clusters randomly, while the clustering-based initialization methods assign individuals into clusters according to some distance criteria.

  \subsection{Identifiability}  \label{identifiabilite}

  \vspace{0.25cm}
  Identifiability of models is necessary to have statistical consistency, which is a minimal requirement for an inference method. The parameters are of two types: $\left(K,\ S\right)$ on one hand, and $\left(\pi,\ \alpha,\ \beta\right)$ on the other hand for a given $\left(K,\ S\right)$. The identifiability of the parameter $\left(K,\ S\right)$ is discussed in sub-section \ref{identifiabilite_KS}. For a given $\left(K,\ S\right)$, the parameter $\beta$ is always identifiable. The identifiability of the parameter $\left(\pi,\ \alpha\right)$ is discussed in sub-section \ref{identifiabilite.theta} using the results of \emph{Elizabeth Allan et al.} \cite{Allman} which is given here in a multi-allelic multilocus genomic data framework.
    \subsubsection{Identifiability of the parameters $\left(K,\ S\right)$}  \label{identifiabilite_KS}
    
    \vspace{0.25cm}
    Let $\mathcal{D}=\bigcup_{\left(K,\ S\right)}\mathcal{M}_{\left(K,\ S\right)}$ be the set of all probability distributions defined by the models $\mathcal{M}_{\left(K,\ S\right)}$ in competition. We assume that the true probability distribution $P_0$ of the observations that we are dealing with is an element of $\mathcal{D}$.
    \vspace{0.50cm}
    
    For a given $P\in\mathcal{D}$, let us define $K\left(P\right)$ and $S\left(P\right)$ as follow.
    \begin{definition}  \label{definition_KS}
      For every $P$ in $\mathcal{D}$, 
      \begin{equation}	\label{def.K}
        K\left(P\right)=\min_{K}\ \left\{K\ :\ P\in\mathcal{M}_{\left(K,\ \cdot\right)}\right\},
      \end{equation}
      \begin{equation}	\label{def.S}
        S\left(P\right)=\min_{S}\ \left\{S\ :\ P\in\mathcal{M}_{\left(\ \cdot,\ S\right)}\right\},
      \end{equation}
      where $\mathcal{M}_{\left(K,\ \cdot\right)}=\bigcup_S\mathcal{M}_{\left(K,\ S\right)}$ and $\mathcal{M}_{\left(\ \cdot,\ S\right)}=\bigcup_K\mathcal{M}_{\left(K,\ S\right)}$.
    \end{definition}
    
    This definition is justified by the following lemmas \ref{Inclusion_KS} and \ref{identifiabilite.KS}.
    \begin{lemma} \label{Inclusion_KS}
      For every $\left(K,\ S\right)$ and $\left(K',\ S'\right)$, if $K\leq K'$ and $S\subseteq S'$, then $\mathcal{M}_{\left(K,\ S\right)}\subseteq\mathcal{M}_{\left(K',\ S'\right)}$. 
    \end{lemma}
    \begin{profLemma}{\ref{Inclusion_KS} : }
      Let $P\in\mathcal{M}_{\left(K,\ S\right)}$ and let $\theta=\left(\pi,\ \alpha,\ \beta\right)\in\Theta_{\left(K,\ S\right)}$ be the parameter defining $P$. Let for instance define $\theta'=\left(\pi',\ \alpha',\ \beta'\right)\in\Theta_{\left(K+1,\ S\right)}$ as follows
      \begin{eqnarray}
        \pi_k'&=&\pi_k,\ k=1,\ldots,\ K-1\nonumber\\
        \pi_K'>0 &\textrm{and}&\pi_{K+1}'>0\textrm{ such that } \pi_K'+\pi_{K+1}'=\pi_K\nonumber\\
        \alpha_{\left(k,\ \cdot,\ \cdot\right)}'&=&\alpha_{\left(k,\ \cdot,\ \cdot\right)},\ k=1,\ldots,\ K\nonumber\\
        \alpha_{\left(K+1,\ \cdot,\ \cdot\right)}'&=&\alpha_{\left(K,\ \cdot,\ \cdot\right)}\nonumber\\
        \beta' &=& \beta\nonumber.
      \end{eqnarray}
      Then we have $\theta'\in\Theta_{\left(K+1,\ S\right)}$ and $P=P_{\left(K+1,\ S\right)}\left(\ \cdot\ |\ \theta'\right)\in\mathcal{M}_{\left(K+1,\ S\right)}$. 

      We have just showed that $\mathcal{M}_{\left(K,\ S\right)}\subseteq\mathcal{M}_{\left(K+1,\ S\right)}$ and there remains to show that $\mathcal{M}_{\left(K,\ S\right)}\subseteq\mathcal{M}_{\left(K,\ S'\right)}$ for every $S$ and $S'$ such that $S\subseteq S'$.
      For such non empty subsets $S$ and $S'$ of available loci, the parameter space $\Theta_{\left(K,\ S\right)}$ can be regarded as a subset of $\Theta_{\left(K,\ S'\right)}$ defined by the following equations :
      \begin{equation}    \label{plonge.S1.S2}
        \alpha_{1,l,\ \cdot}=\ldots =\alpha_{K,l,\ \cdot}\ \forall l\in S'\smallsetminus S.
      \end{equation}
    \end{profLemma}
    \vspace{0.25cm}
    
    \begin{lemma}	\label{identifiabilite.KS}
      For every $K_1,\ K_2\in\left\{1,\ldots,\ K_{\max}\right\}$ and $S_1,\ S_2\in\mathcal{P}^{*}\left(L\right)$, we have $\mathcal{M}_{\left(K_1,\ S_1\right)}\cap\mathcal{M}_{\left(K_2,\ S_2\right)}=\mathcal{M}_{\left(K_1\wedge K_2,\ S_1\cap S_2\right)}$, where $K_1\wedge K_2 = \min \{K_1 ; K_2\}$.
    \end{lemma}
    \begin{profLemma}{\ref{identifiabilite.KS} : }
      Let $P$ be a probability distribution in $\mathcal{M}_{\left(K_1,\ S_1\right)}\cap\mathcal{M}_{\left(K_2,\ S_2\right)}$. Then for every $x$ in $\mathcal{X}$, $P\left(x\right)$ is given by the following two equations.
      \begin{eqnarray}   \label{eq.S1.S2}
          P\left(x\right) &=& \left[\sum_{k=1}^{K_1}\pi_k^1\prod_{l\in S_1}P\left(x^l|\ \left(\alpha_{k,\ l,\ \cdot}^1\right)\right)\right]\times \prod_{l\in S_1^c}P\left(x^l|\ \left(\beta_{l,\ \cdot}^1\right)\right),  \label{eq.K1.S1}\\
          P\left(x\right) &=& \left[\sum_{k=1}^{K_2}\pi_k^2\prod_{l\in S_2}P\left(x^l|\ \left(\alpha_{k,\ l,\  \cdot}^2\right)\right)\right]\times \prod_{l\in S_2^c}P\left(x^l|\ \left(\beta_{l,\ \cdot}^2\right)\right).  \label{eq.K2.S2}
      \end{eqnarray}
      Assume without lost of generality that $K_1\leq K_2$ and denote $A:=S_1\smallsetminus (S_1 \cap S_2 )$, $B:=S_2\smallsetminus (S_1 \cap S_2 )$ and $C=L\smallsetminus S_1\cup S_2$. Using equation (\ref{eq.K1.S1}), the marginal probability distribution of the sub-vector $x^{S_2}:=\left(x^l\right)_{l\in S_2}$ is given by
      \begin{equation}
        P\left(x^{S_2}\right)=\left[\sum_{k=1}^{K_1}\pi_k^1\prod_{l\in S_1\cap S_2}P\left(x^l|\ \left(\alpha_{k,\ l,\ \cdot}^1\right)\right)\right]\times \prod_{l\in B}P\left(x^l|\ \left(\beta_{l,\ \cdot}^1\right)\right),
      \end{equation}
which using equation (\ref{eq.K2.S2}) becomes
      \begin{align}	\label{eq.K.S}
        P\left(x\right)=&\left[\sum_{k=1}^{K_1}\pi_k^1\prod_{l\in S_1\cap S_2}P\left(x^l|\ \left(\alpha_{k,\ l,\ \cdot}^1\right)\right)\right]\times\prod_{l\in B}P\left(x^l|\ \left(\beta_{l,\ \cdot}^1\right)\right)\nonumber\\
        & \times\prod_{l\in A\cup C}P\left(x^l|\ \left(\beta_{l,\ \cdot}^2\right)\right)\nonumber\\
        =&\left[\sum_{k=1}^{K_1}\pi_k^1\prod_{l\in S_1\cap S_2}P\left(x^l|\ \left(\alpha_{k,\ l,\ \cdot}^1\right)\right)\right]\times \prod_{l\in A\cup B\cup C}P\left(x^l|\ \left(\beta_{l,\ \cdot}^3\right)\right),\nonumber
      \end{align}
      which implies that $P\in\mathcal{M}_{\left(K_1\wedge K_2,\ S_1\cap S_2\right)}$
    \end{profLemma}
    \vspace{0.50cm}
    
    Obviously, by definition \ref{definition_KS}, for every $P_1$ and $P_2$ in $\mathcal{D}$, 
    \begin{equation}
      \left(P_1=P_2\right)\Longrightarrow \left[K\left(P_1\right)=K\left(P_2\right)\textrm{ and }S\left(P_1\right)=S\left(P_2\right)\right].
    \end{equation}
    We will denote $\left(K_0,\ S_0\right):=\left(K\left(P_0\right),\ S\left(P_0\right)\right)$, and this definition is well compatible with the use of the BIC criterion which aims at selecting the smallest model dimension in statistical
    adjustment.

    \subsubsection{Identifiability of parameter $\gamma=\left(\pi,\ \alpha\right)$ in the model $\mathcal{M}_{\left(K,\ S\right)}$} \label{identifiabilite.theta}
    
    \vspace{0.25cm}
    The classical definition of an identifiable model $\mathcal{M}_{\left(K,\ S\right)}$ of probability distributions requires that for any two different parameter values $\theta$ and $\theta'$ in parameter space $\Theta_{\left(K,\ S\right)}$, the corresponding probability distributions $P_{\left(K,\ S\right)}\left(\ \cdot\ |\ \theta\right)$ and $P_{\left(K,\ S\right)}\left(\ \cdot\ |\ \theta'\right)$ be different. This is to require injectivity of the parameterization map $\Psi$ for this model, which is defined by $\Psi\left(\theta\right)=P_{\left(K,\ S\right)}\left(\ \cdot\ |\ \theta\right)$.
    \vspace{0.25cm}
    
    In our context of finite mixtures, the above map will not strictly be injective because the latent classes can be freely relabeled without changing the distribution underlining the observations. This is known as 'label swapping'. In such a case, the above map is always at least $K!$-to-one.
    \vspace{0.25cm}
    
    If the model is identifiable up to label swipping, then the number of independent parameters is at most equal to the number of distinct genotypes :
    \begin{equation}    
      K-1 + K{\sum}_{l\in S}\left(A_l-1\right)\leq\prod_{l\in S}\left(\left(\begin{array}{c} 2\\ A_l\end{array}
                                                                                                    \right)+A_l\right)-1 .
    \end{equation}
    Despite that this condition is not sufficient, it gives  an upper bound on $K_{\max}=\max_{S}K\left(S\right)$ of the number of populations where
    \begin{equation}	\label{K.max.KS}
      K\left(S\right):=\frac{\prod_{l\in S}\frac{A_l\left(A_l+1\right)}{2}}{1+\sum_{l\in S}\left(A_l-1\right)}.
    \end{equation}
    We use that upper bound to define the collection of models in competition given by equation (\ref{collection_modeles}).
    \vspace{0.25cm}
    
    Assume that the frequencies of the distinct observed genotypes are the parameters of interest. For a given $K$ and $S$, we refer to the finite mixture model (\ref{Melange}) as the $K$-class, $|S|$-feature model, with state space $\prod_{l\in S}\left\{1,\ldots,\ G_l\right\}$, as $\mathcal{M}\left(K\ ;\ \left(G_l\right)_{l\in S}\right)$, where $G_l=\frac{A_l\left(A_l+1\right)}{2}$ is the number of distinct observed genotypes at locus $l$ and $|S|$ the cardinality of $S$. 
    \vspace{0.25cm}
    
    \emph{Elizabeth S. Allman et al.} (2008) \cite{Allman} has proved that finite mixtures of multinomial distributions are \emph{generically} identifiable. In the case of parametric setting, 'generic' means that the set of points for which identifiability does not hold has zero-measure. Here is the result  of \emph{Elizabeth S. et al.} relevant to our setting.
    \begin{theorem} \label{theorem.Identifiabilite}
      Consider the model $\mathcal{M}\left(K\ ;\ \left(G_l\right)_{l\in S}\right)$ where $|S|\geq 3$. Assume there exists a tripartition of the set $S$ into three disjoint non-empty subsets $S_1$, $S_2$ and $S_3$, such that if $\mathcal{G}_i=\prod_{l\in S_i}G_l$, then 
      \begin{equation}
        \min\left(K,\ \mathcal{G}_1\right)+\min\left(K,\ \mathcal{G}_2\right)+\min\left(K,\ \mathcal{G}_3\right)\geq 2\cdot K+2.
      \end{equation}
      Then the model is generically idenfiable, up to label swapping. Moreover, the statement remains valid when the proportions of the groups $\left\{\pi_k\right\}_{k=1,\ldots,\ K}$ are held fixed and positive.
    \end{theorem}
    
    This result implies that one needs a minimum of genetic variability to guarantee the identifiability of the models in competition. For example, it will be difficult to detect $4$ subpopulations with $3$ biallelic loci such as Single Nucleotide Polymorphims (SNP).
\section{Consistency} \label{consistency}

  \vspace{0.25cm}
  In this section, it is proved that the probability of selecting the true model $\left(K_0,\ S_0\right)$ by maximizing criterion (\ref{BIC}) tends to $1$ as $n\rightarrow\infty$ under the following single assumption:
  \begin{equation}  \label{hypothese}
    \left(H\right)\ :\ \forall u\in\mathcal{X},\ P_0\left(u\right)>0,
  \end{equation}
  where $\mathcal{X}$ is the set of distinct genotypes defined by the observed alleles, and $P_0$ the true probability distribution of the observations. Assumption $\left(H\right)$ is realistic because our method is proposed for experiments in which only observed alleles are considered.   
  \begin{theorem} \label{theorem1}
    Under assumption $\left(H\right)$,
     $$ \lim_{n\rightarrow\infty}P_{0}\left[\left(\widehat{K}_n,\ \widehat{S}_n\right)=\left(K_0,\ S_0\right)\right]=1.$$
  \end{theorem}
  \begin{profTheorem}{\ref{theorem1} : \\}
  
    We need to prove that $\lim_{n\rightarrow\infty}P_{0}\left[\left(\widehat{K}_n,\ \widehat{S}_n\right)\neq\left(K_0,\ S_0\right)\right]=0$.
    $$
      P_{0}\left[\left(\widehat{K}_n,\ \widehat{S}_n\right)\neq\left(K_0,\ S_0\right)\right]\leq\sum_{\left(K,\ S\right)\neq\left(K_0,\ S_0\right)}P_{0}\left[\left(\widehat{K}_n,\ \widehat{S}_n\right)=\left(K,\ S\right)\right],
    $$
 so that since the number of possible $\left(K,\ S\right)$ is finite, the theorem is proved if for every $\left(K,\ S\right)\neq\left(K_0,\ S_0\right)$, $\lim_{n\rightarrow\infty}P_{0}\left[\left(\widehat{K}_n,\ \widehat{S}_n\right)=\left(K,\ S\right)\right]= 0$.
    
    \vspace{0.25cm}
    Let $\left(K,\ S\right)\in\left\{1,\ldots,\ K_{\max}\right\}\times\mathcal{P}^{*}\left(L\right)$ such that $\left(K,\ S\right)\neq\left(K_0,\ S_0\right)$. We have 
    \begin{align}  \label{theorem1_1}
     &P\left[\left(\widehat{K}_n,\ \widehat{S}_n\right)=\left(K,\ S\right)\right]\leq P\Bigg[2\sup_{\theta\in\Theta_{\left(K,\ S\right)}}\ell_n\bigg(P_{\left(K,\ S\right)}\left(\ \cdot\ |\ \theta\right)\bigg)\nonumber \\
     &-2\sup_{\theta\in\Theta_{\left(K_0,\ S_0\right)}}\ell_n\bigg(P_{\left(K_0,\ S_0\right)}\left(\ \cdot\ |\ \theta\right)\bigg)>\bigg(d_{\left(K,\ S\right)}-d_{\left(K_0,\ S_0\right)}\bigg)\ln n\Bigg]
    \end{align}
    where $d_{\left(K,\ S\right)}$ is the number of independent parameters of model $\mathcal{M}_{\left(K,\ S\right)}$, and 
    $$
      \ell_n\left(P\right)=\sum_{u\in\mathcal{X}}n_u\ln P\left(u\right)
    $$ 
  is  the log-likelihood. Here $n_u$ is the number of individuals in the sample with genotype $u$. Two cases are considered  : $P_0\in\mathcal{M}_{\left(K,\ S\right)}$ and $P_0\notin\mathcal{M}_{\left(K,\ S\right)}$.
    \vspace{0.25cm}
    
    $\bullet$ {\textbf{Case 1 :} $P_0\in\mathcal{M}_{\left(K,\ S\right)}$. \\}
    
    Let $\mathcal{D}\left(\mathcal{X}\right)$ denote the set of all probability distributions on the set $\mathcal{X}$ of distinct observed genotypes. Since $\mathcal{M}_{\left(K,\ S\right)}\subset \mathcal{D}\left(\mathcal{X}\right)$,
    $$
      \ell_n\left(P_0\right)\leq\sup_{\theta\in\Theta_{\left(K,\ S\right)}}\ell_n\bigg(P_{\left(K,\ S\right)}\left(\ \cdot\ |\ \theta\right)\bigg)\leq
 \sup_{P\in\mathcal{D}\left(\mathcal{X}\right)}\ell_n\left( P\right)    ,
    $$
so that
$$
      0\leq\sup_{\theta\in\Theta_{\left(K,\ S\right)}}\ell_n\bigg(P_{\left(K,\ S\right)}\left(\ \cdot\ |\ \theta\right)\bigg)-\ell_n\left(P_0\right)
\leq
 \sup_{P\in\mathcal{D}\left(\mathcal{X}\right)}\ell_n\left( P\right)-\ell_{n }\left( P_{0}\right).
$$
But it is well known that $2 \sup_{P\in\mathcal{D}\left(\mathcal{X}\right)}\ell_n\left( P\right)-2\ell_n\left(P_0\right)$
converges in distribution to a chi-square variable with 
$|\mathcal{X}|-1$ numbers of freedom, where $|\mathcal{X}|$ denote the cardinality of $\mathcal{X}$. Also, if $P_0\in\mathcal{M}_{\left(K,\ S\right)}$ and $\left(K,\ S\right)\neq\left(K_0,\ S_0\right)$, $d_{\left(K,\ S\right)}-d_{\left(K_0,\ S_0\right)}>0$. Thus in this case
\begin{align*}
 \lim_{n\rightarrow\infty}P\Bigg[2\sup_{\theta\in\Theta_{\left(K,\ S\right)}}\ell_n\bigg(P_{\left(K,\ S\right)}\left(\ \cdot\ |\ \theta\right)\bigg)
 - & 2\sup_{\theta\in\Theta_{\left(K_0,\ S_0\right)}}\ell_n\bigg(P_{\left(K_0,\ S_0\right)}\left(\ \cdot\ |\ \theta\right)\bigg)\\
   > &  \bigg(d_{\left(K,\ S\right)}-d_{\left(K_0,\ S_0\right)}\bigg)\ln n\Bigg]=0.
\end{align*}
\vspace{0.25cm}

    $\bullet$ {\textbf{Case 2 :} $P_0\notin\mathcal{M}_{\left(K,\ S\right)}$\\}%

    For every $\delta>0$, let 
    $$\Theta_{\left(K,\ S\right)}^{\delta}=\left\{\theta\in\Theta_{\left(K,\ S\right)}\ :\ \forall x\in \mathcal{X},\ P_{\left(K,\ S\right)}\left(x|\ \theta\right)\geq \delta\right\}.$$
    The key point is the following:
  \begin{proposition}	\label{proposition2}
    Under assumption $\left(H\right)$, there exists a real $\delta>0$ such that for every $\left(K,\ S\right)$, 
    \begin{equation}
      \sup_{\theta\in\Theta_{\left(K,\ S\right)}}\frac{1}{n}\ell_n\bigg(P_{\left(K,\ S\right)}\left(\ \cdot\ |\ \theta\right)\bigg)=\sup_{\theta\in\Theta_{\left(K,\ S\right)}^{\delta}}\frac{1}{n}\ell_n\bigg(P_{\left(K,\ S\right)}\left(\ \cdot\ |\ \theta\right)\bigg)+o_{P_0}\left(1\right).
    \end{equation}
  \end{proposition}
\begin{profProposition}{\ref{proposition2} : \\}
    \begin{align}  \label{eeee}
      \frac{1}{n}\ell_n\bigg(P_{\left(K,\ S\right)}\left(\ \cdot\ |\ \theta\right)\bigg)&=\sum_{u\in\mathcal{X}}\frac{n_u}{n}\ln P_{\left(K,\ S\right)}\left(u|\ \theta\right)\nonumber\\
      &= \sum_{u\in\mathcal{X}}\bigg[P_0\left(u\right)+o_{P_0}\left(1\right)\bigg]\times\ln P_{\left(K,\ S\right)}\left(u|\ \theta\right).
    \end{align}
    The set of $\widetilde{\delta}>0$ such that $\Theta_{\left(K,\ S\right)}^{\widetilde{\delta}}\neq\emptyset$. Let $\widetilde{\delta}>0$ be such a real and $\widetilde{\theta}$ an element of $\Theta_{\left(K,\ S\right)}^{\widetilde{\delta}}$. Since for any $u$, $P_{0}(u)>0$, using (\ref{eeee}),  
 \begin{equation}	\label{lemme.Theta.delta.2}
      \frac{1}{n}\ell_n\bigg(P_{\left(K,\ S\right)}\left(\ \cdot\ |\ \widetilde{\theta}\right)\bigg)\geq\sum_{x\in\mathcal{X}}P_0\left(x\right)\ln \widetilde{\delta}+o_{P_0}\left(1\right)
 = \ln\widetilde{\delta}+o_{P_0}\left(1\right).
    \end{equation}
    Let $\delta$ be a real such that $0<\delta<\widetilde{\delta}^{\frac{1}{\inf_{u\in\mathcal{X}}P_0\left(u\right)}}$ and $\delta\leq\inf_{u\in\mathcal{X}}P_0\left(u\right)$. Remark that $0<\inf_{u\in\mathcal{X}}P_0\left(u\right)\leq 1$ and $0<\widetilde{\delta}<1$ imply
    $
      \frac{1}{\inf_{\mathcal{X}}P_0\left(x\right)}\ln \widetilde{\delta}\leq\ln \widetilde{\delta}, 
    $
so that    $0<\delta<\widetilde{\delta}^{\frac{1}{\inf_{\mathcal{X}}P_0\left(x\right)}}\leq\widetilde{\delta}$.

Then $\Theta_{\left(K,\ S\right)}^{\widetilde{\delta}}\subset \Theta_{\left(K,\ S\right)}^{\delta}$, and thus 
$$
      \sup_{\theta\in\Theta_{\left(K,\ S\right)}^{\delta}}\frac{1}{n}\ell_n\bigg(P_{\left(K,\ S\right)}\left(\ \cdot\ |\ \theta\right)\bigg)\geq \sup_{\theta\in\Theta_{\left(K,\ S\right)}^{\widetilde{\delta}}}\frac{1}{n}\ell_n\bigg(P_{\left(K,\ S\right)}\left(\ \cdot\ |\ \theta\right)\bigg).
$$

If now $\theta\in\Theta_{\left(K,\ S\right)}\smallsetminus\Theta_{\left(K,\ S\right)}^{\delta}$, then there exists a genotype $u_{\delta}\in\mathcal{X}$ such that $P_{\left(K,\ S\right)}\left(u_{\delta}|\ \theta\right)<\delta$. In such a case
    \begin{align*}
      \frac{1}{n}\ell_n\bigg(P_{\left(K,\ S\right)}\left(\ \cdot\ |\ \theta\right)\bigg) &\leq \inf_{\mathcal{X}}P_0\left(u\right)\ln \delta +o_{P_0}\left(1\right)\\
      &\leq \inf_{\mathcal{X}}P_0\left(u\right)\ln\widetilde{\delta}^{\frac{1}{\inf_{\mathcal{X}}P_0\left(x\right)}}+o_{P_0}\left(1\right)\ = \ln\widetilde{\delta}+o_{P_0}\left(1\right)\\
      &\leq \sup_{\theta\in\Theta_{\left(K,\ S\right)}^{\widetilde{\delta}}}\frac{1}{n}\ell_n\bigg(P_{\left(K,\ S\right)}\left(\ \cdot\ |\ \theta\right)\bigg)+o_{P_0}\left(1\right)\\
      &\leq \sup_{\theta\in\Theta_{\left(K,\ S\right)}^{\delta}}\frac{1}{n}\ell_n\bigg(P_{\left(K,\ S\right)}\left(\ \cdot\ |\ \theta\right)\bigg)+o_{P_0}\left(1\right)
    \end{align*}
    Thus, 
 $$ \sup_{\theta\in\Theta_{\left(K,\ S\right)}}\frac{1}{n}\ell_n\bigg(P_{\left(K,\ S\right)}\left(\ \cdot\ |\ \theta\right)\bigg) = \sup_{\theta\in\Theta_{\left(K,\ S\right)}^{\delta}}\frac{1}{n}\ell_n\bigg(P_{\left(K,\ S\right)}\left(\ \cdot\ |\ \theta\right)\bigg)+o_{P_0}\left(1\right)
 $$
    which is the desired result.
\end{profProposition}\\
\vspace{0.25cm}

Now, the set of functions $\{\ln P_{\left(K,\ S\right)}\left(\ \cdot\ |\ \theta\right),\theta \in \Theta_{\left(K,\ S\right)}^{\delta} \}$ is obviously Glivenko-Cantelli, so that
$$
   \sup_{\theta\in\Theta_{\left(K,\ S\right)}^{\delta}}\frac{1}{n}\ell_n\bigg(P_{\left(K,\ S\right)}\left(\ \cdot\ |\ \theta\right)\bigg)=\sup_{\theta\in\Theta_{\left(K,\ S\right)}^{\delta}}\ E_{P_{0}}\bigg[\ln P_{\left(K,\ S\right)}\left(U|\ \theta\right)\bigg]+o_{P_0}\left(1\right),
$$  
and Proposition \ref{proposition2} yields, for any $(K,S)$, 
$$
      \sup_{\theta\in\Theta_{\left(K,\ S\right)}}\frac{1}{n}\ell_n\bigg(P_{\left(K,\ S\right)}\left(\ \cdot\ |\ \theta\right)\bigg)=\sup_{\theta\in\Theta_{\left(K,\ S\right)}^{\delta}}\ E_{P_{0}}\bigg[\ln P_{\left(K,\ S\right)}\left(U|\ \theta\right)\bigg]+o_{P_0}\left(1\right).
$$
Also,
$$
\sup_{\theta\in\Theta_{\left(K_{0},\ S_{0}\right)}^{\delta}}\ E_{P_{0}}\bigg[\ln P_{\left(K_{0},\ S_{0}\right)}\left(U|\ \theta\right)\bigg]
=E_{P_{0}} \ln P_{0}\left(U\right),
$$
since $P_0\in\mathcal{M}_{\left(K_0,\ S_0\right)}$ and $P_0\left(u\right)\geq\delta$ $\forall u\in\mathcal{X}$.
Thus
\begin{align*}
  \frac{1}{n}\sup_{\theta\in\Theta_{\left(K,\ S\right)}}&\ell_n\bigg(P_{\left(K,\ S\right)}\left(\ \cdot\ |\ \theta\right)\bigg)
 -\frac{1}{n}\sup_{\theta\in\Theta_{\left(K_0,\ S_0\right)}}\ell_n\bigg(P_{\left(K_0,\ S_0\right)}\left(\ \cdot\ |\ \theta\right)\bigg) =\\
  & -\inf_{\theta\in\Theta_{\left(K,\ S\right)}^{\delta}} 
E_{P_{0}}\bigg[\ln P_{0}\left(U\right)-\ln P_{\left(K,\ S\right)}\left(U|\ \theta\right)\bigg]
+o_{P_0}\left(1\right).
\end{align*}
But on the compact set $\Theta_{\left(K,\ S\right)}^{\delta}$, the function $\theta \mapsto 
E_{P_{0}}\bigg[\ln P_{0}\left(U\right)-\ln P_{\left(K,\ S\right)}\left(U|\ \theta\right)\bigg]$ is continuous
and attains its infimum at a point $\overline{\theta}$. But since $P_0\notin\mathcal{M}_{\left(K,\ S\right)}$,
$P_{0}\left(\cdot \right) \neq P_{\left(K,\ S\right)}\left(\cdot|\ \overline{\theta}\right)$, and
$$
E_{P_{0}}\bigg[\ln P_{0}\left(U\right)-\ln P_{\left(K,\ S\right)}\left(U|\ \overline{\theta}\right)\bigg]>0.
$$
 Noticing that $\lim_{n\rightarrow \infty}\frac{(d_{(K,S)}-d_{(K_{0},S_{0})})\ln n}{n}=0$, one gets
\begin{align*}
\lim_{n\rightarrow +\infty}
P\Bigg[2\sup_{\theta\in\Theta_{\left(K,\ S\right)}}\ell_n\bigg(P_{\left(K,\ S\right)}\left(\ \cdot\ |\ \theta\right)\bigg)
 &-2\sup_{\theta\in\Theta_{\left(K_0,\ S_0\right)}}\ell_n\bigg(P_{\left(K_0,\ S_0\right)}\left(\ \cdot\ |\ \theta\right)\bigg)\\
 & > \bigg(d_{\left(K,\ S\right)}-d_{\left(K_0,\ S_0\right)}\bigg)\ln n\Bigg]=0.
\end{align*}
\end{profTheorem}
\section{Simulation examples} \label{simulation}

  \textbf{MixMoGenD} has been implemented using $C++$ and $C$ programming languages. The main goal of the simulation examples was to confirm in practice the consistency of the loci selection procedure and to highliht its benifits. Before that, preliminary simulations were conducted to regulate certain known problems of the EM algorithm, in particular convergence towards the maximum likelihood and the low speed of convergence in certain cases. In fact, the EM algorithm converges almost always towards a local maximum under certain conditions of regularity. Thus it is not certain whether the algorithm converges towards a local or global maximum when there are several maxima. To reduce the dependence of the convergence point to the initial parameter of the algorithm, we opt for the strategy of at least $50$ initial parameters, and the maximum likelihood estimate is the one maximasing the likelihood. For each initial parameter, we stop the EM algorithm when the difference between two consecutive likelihoods of the complete data is less than a certain positive real $\varepsilon>0$ to be chosen by the user.
  \subsection{Consistency of the selection procedure}
  
\begin{figure}[htbp]  
     \centering{\includegraphics[width=3.30in,height=3.30in]{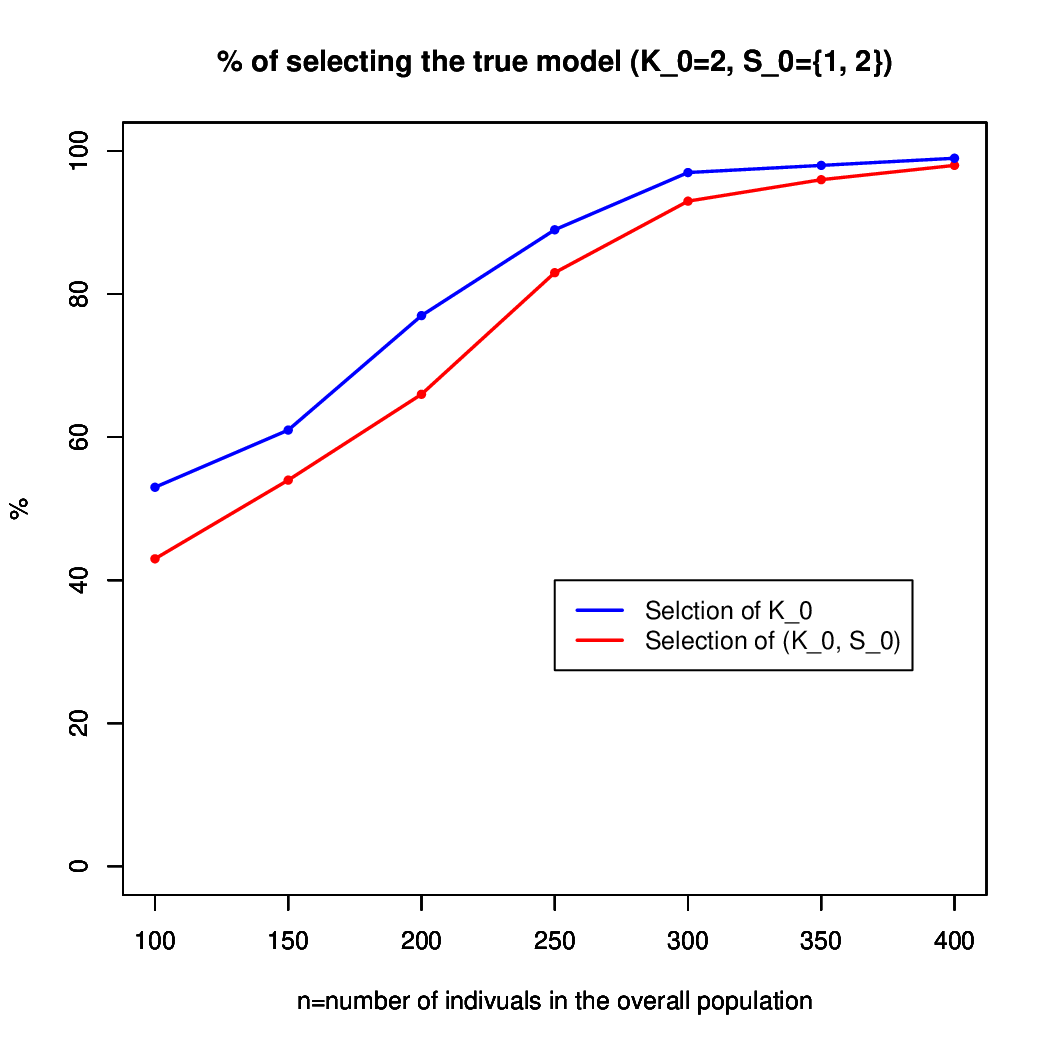}}
     \caption{\label{Fig1}{\small{\% of selecting the true model vs number of individuals}}}
 \end{figure}

  The goal here was to see how the increase of the size of the sample improves the capacity of our clustering method to select the true model $\mathcal{M}_{\left(K_0,\ S_0\right)}$. An interface between \textbf{MixMoGenD} and \textbf{R} was created for these simulations. In these experiments, we started  with $n=100$ individuals, and gradually increased this number to $400$ by a step of $50$. We assumed $K_0=2$ populations, $L=4$ loci with $2$ alleles by locus, $S_0$ with cardinality $|S_0|=2$. For each value $n$ of the sample size, $100$ data sets were generated. The parameters of simulation are given in Table \ref{Tab1}. The figure \ref{Fig1} shoes that \textbf{MixMoGenD} consistently identify the true model as $n\rightarrow\infty$. Other simulated data with $K_0=3$, $|S_0|=4$ and $|S_0^c|=2$ confirmed these results. These results confirm the theoretical result on the consistency that we showed in Section \ref{consistency}.
  \subsection{Benefits of the selection procedure}
  
  Two series of simulations were conducted to highlight the importance of the loci selection procedure. First, we independently generated $100$ data sets, each of them contained $1\ 000$ individuals. We assumed $K_0=3$ populations with the proportions given by $\pi=\left(0.20,\ 0.30,\ 0.50\right)$, $L=6$ loci with the numbers of alleles given by $\left(3,\ 4,\ 3,\ 3,\ 3,\ 4\right)$, $S_0=\left\{1,\ 2,\ 3,\ 4\right\}$ and allele frequencies given in Table \ref{Tab2}. Using all the $6$ loci, the true model was selected $39$ times against $61$ for the model with $\widehat{K}_n=2$. When including the selection procedure, \textbf{MixMoGenD} selected the true model $\left(K_0,\ S_0\right)$ $90$ times against $10$ for $\left(K,\ S\right)=\left(2,\ S_0\right)$. It appears that the number of populations is under estimated when considering all available loci as relevant for clustering. 
  \vspace{0.25cm}
  
  To confirm this result, a second series of simulations with more variability was conducted. In these simulations, each of the data sets consisted of $1\ 000$ individuals structured into $5$ subpopulations of equal proportions. We assumed $L=10$ loci each with $10$ alleles, and two different cardinalities for $S_0$: $8$ and $6$. Instead of choosing manualy the allelic frequencies, we adopt the following strategy. For the loci in $S_0$, we first use the program \textbf{EASYPOP} \cite{Balloux2001} to simulate some data sets at some levels of $F_{ST}$ between $0.03$ and $0.04$. Second, we estimated the allelic frequencies of the loci in $S_0$ by EM algorithm. And thirdly, we used these estimates and uniform probability distribution on loci in $S_0^c$ to simulate the data sets with \textbf{R} program \cite{R}. Sample project files and their simulation parameters are available on \url{http://www.math.u-psud/~toussile}.
    \subsubsection*{\textbf{Results}}
    
    As expected, the results in the Table \ref{Tab3} show that the integrated loci selection procedure significantly improves the inference on the number $K$ of subpopulations and the quality of the prediction. The benefit of the selection is more important with the increase of cardinality of the subset $S_0^c$ of loci that are not relevant for clustering. The important result is that for these simulations, \textbf{MixMoGenD} perfectly selected the true subset $S_0$ of relevant loci for clustering. For each data set for which $\widehat{K}_n<K_0$, we calculated the square matrix of the pairwise $F_{ST}$ between individuals sampled using the function \emph{Fstat} of the package \emph{Geneland} \cite{Guillot2005} of \textbf{R} program. We observed that there exists a threshold $F_{ST_{\max}}$ of pairwise $F_{ST}$ for which two subpopulations with $F_{ST}<F_{ST_{\max}}$ are clustered together. This threshold are approximately $0.027$ on the simulated data sets we used. The more striking example is the data set $5$ in Table \ref{Tab3} (d). The square matrix of pairwise $F_{ST}$ is given in Table \ref{Tab4}. The $F_{ST}$ between population $4$ and the others are all $<0.026$. On this data set, MixMoGenD produces $4$ clusters and we observed that Pop4 was uniformly distributed in the $4$ clusters.
\section{Discussion}
\label{Discussion}

  We believe that \textbf{MixMoGenD} will be useful for two main reasons. First, like \textbf{FASTRUCT}, \textbf{MixMoGenD} is based on the EM algorithm, so that both share certain qualities, particularly they are faster than their counterparts based on a bayesian approach \cite{Francois2006}. 
  
  The key point of our proposed method is that it is combined with a loci selection procedure. That is the main reason for which our method will be very useful, and it is our main contribution. In fact, the results obtained on simulated data show how the selection procedure improves significantly the inference on the number $K$ of subpopulations and the prediction capacity. In addition, due to the explosion of genomic projects, data sets are becoming increasingly large. The space of the models in competion can then be very large. Then an exhaustive research of an optimum model is very painful in most situation and could not be achieved by methods based on MCMC algorithm as mentioned by \emph{O. Francois et al.} (2006) \cite{Chen2006}. Thus methods like frequentist likelihood methods using EM algorithm will then become useful because they require much shorter computations than the methods based on MCMC algorithm. For example, \emph{E. K. Latch} (2005) \cite{Latch2006} reported that a data set with $5$ subpopulations, $100$ individuals in each subpopulation, $10$ loci and $10$ alleles by locus take approximately $3\ h$ to run without loci selection on \textbf{STRUCTURE} \cite{Pritchard2000}, and $30\ h$ on \textbf{PARTITION} (all times provided are appropriate for a computer with a $2.2\ GHz$ Celeron processor and $512\ MB$ of RAM). For such data sets, \textbf{MixMoGenD} and its selection procedure take approximately $2\ h\ 30$ to run. This was made possible thanks to the Backward-Stepwise algorithm, which enabled us to avoid an exhaustive research of the optimum model among all the models in competition.

\section*{Acknowledgements}

  This work was supported by a doctoral fellowship from the "Institut de Recherche pour le D\'eveloppement" (IRD), in the framework of the research unit UR06 of this insititut. We are gratefull to \emph{Isabelle Morlais} for the explanations of the biological concepts we needed, and to \emph{Henri GWET} for his constructive suggestions and his encouragement.



\nocite{}
\bibliographystyle{plain}
\bibliography{mabiblio}


\newpage
\begin{table}[htbp]
 \centering
 \begin{tabular}{cccc}
  \hline
  Locus & Allele & Pop1 & Pop2 \\
  \hline\hline
  $1$ & $1$ & $0.70$ & $0.25$ \\
      & $2$ & $0.30$ & $0.75$ \\
  \hline
  $2$ & $1$ & $0.35$ & $0.70$ \\
      & $2$ & $0.65$ & $0.30$ \\
  \hline
 \end{tabular}
 \begin{tabular}{cccc}
  \hline
  Locus & Allele & Pop1 & Pop2 \\
  \hline\hline
  $3$ & $1$ & $0.85$ & $0.85$ \\
      & $2$ & $0.15$ & $0.15$ \\
  \hline
  $4$ & $1$ & $0.50$ & $0.50$ \\
      & $2$ & $0.50$ & $0.50$ \\
  \hline
 \end{tabular}
 \caption{\label{Tab1}{\small{Parameters of simulated data to show the consistency of the selection procedure. $K_0=2$, $S_0=\left\{1,\ 2\right\}$, $\pi=\left(0.30,\ 0.70\right)$.}}}
\end{table}
\begin{table}[htbp]
 \centering
 \begin{tabular}{ccccc}
  \hline
  L & Allele & Pop1 & Pop2 & Pop3 \\
  \hline\hline
  $1$ & $1$ & $0.20$ & $0.40$ & $0.50$ \\
      & $2$ & $0.30$ & $0.40$ & $0.20$ \\
      & $3$ & $0.50$ & $0.20$ & $0.30$ \\
  \hline
  $2$ & $1$ & $0.20$ & $0.40$ & $0.50$ \\
      & $2$ & $0.20$ & $0.40$ & $0.10$ \\
      & $3$ & $0.40$ & $0.10$ & $0.10$ \\
      & $4$ & $0.20$ & $0.10$ & $0.30$ \\
  \hline
  $3$ & $1$ & $0.15$ & $0.25$ & $0.50$ \\
      & $2$ & $0.25$ & $0.25$ & $0.10$ \\
      & $3$ & $0.60$ & $0.50$ & $0.40$ \\
  \hline
 \end{tabular}
 \begin{tabular}{ccccc}
  \hline
  L & Allele & Pop1 & Pop2 & Pop3 \\
  \hline\hline
  $4$ & $1$ & $0.30$ & $0.40$ & $0.65$ \\
      & $2$ & $0.60$ & $0.40$ & $0.15$ \\
      & $3$ & $0.10$ & $0.20$ & $0.20$ \\
  \hline
  $5$ & $1$ & $0.25$ & $0.25$ & $0.25$ \\
      & $2$ & $0.30$ & $0.30$ & $0.30$ \\
      & $3$ & $0.25$ & $0.25$ & $0.25$ \\
      & $4$ & $0.20$ & $0.20$ & $0.20$ \\
  \hline
  $6$ & $1$ & $0.40$ & $0.40$ & $0.40$ \\
      & $2$ & $0.30$ & $0.30$ & $0.30$ \\
      & $3$ & $0.30$ & $0.30$ & $0.30$ \\
  \hline
 \end{tabular}
  \caption{\label{Tab2}{\small{Parameters of simulated data to show the benefit of the selection procedure: $K_0=3$, $\pi=\left(0.20,\ 0.30,\ 0.50\right)$, $S_0=\left\{1,\ 2,\ 3,\ 4\right\}$. L = locus, Pop=Population}}}
\end{table}
\begin{table}[htbp]
 \centering
 \begin{tabular}{ccccc}
  \hline
  Data & $\widehat{K}_n^s$ & $\%$ MA & $\widehat{K}_n$ & $\%$ MA \\
  \hline\hline
  $1$ & $4$ & & $3$ & \\
  $2$ & $5$ & $09.10$ & $3$ & \\
  $3$ & $3$ & & $3$ & \\
  $4$ & $3$ & & $3$ & \\
  $5$ & $5$ & $12.40$ & $3$ & \\
  $6$ & $4$ & & $4$ & \\
  $7$ & $3$ & & $3$ & \\
  $8$ & $3$ & & $3$ & \\
  $9$ & $5$ & $11.80$ & $3$ & \\
  $10$ & $3$ & & $3$ & \\
  \hline
  (a)
 \end{tabular}
 \begin{tabular}{ccccc}
  \hline
  Data & $\widehat{K}_n^s$ & $\%$ MA & $\widehat{K}_n$ & $\%$ MA \\
  \hline\hline
  $1$ & $5$ & $08.00$ & $5$ & $08.80$ \\
  $2$ & $5$ & $08.90$ & $4$ & \\
  $3$ & $5$ & $10.40$ & $5$ & $11.40$ \\
  $4$ & $5$ & $10.20$ & $5$ & $10.50$ \\
  $5$ & $5$ & $08.80$ & $5$ & $09.30$ \\
  $6$ & $5$ & $10.20$ & $5$ & $10.30$ \\
  $7$ & $5$ & $09.10$ & $4$ & \\
  $8$ & $5$ & $07.60$ & $5$ & $08.50$ \\
  $9$ & $5$ & $09.50$ & $4$ & \\
  $10$ & $5$ & $10.30$ & $5$ & $10.90$ \\
  \hline
  (b)
 \end{tabular}
 \\
 \begin{tabular}{ccccc}
  \hline
  Data & $\widehat{K}_n^s$ & $\%$ MA & $\widehat{K}_n$ & $\%$ MA \\
  \hline\hline
  $1$ & $5$ & $07.50$ & $5$ & $07.10$ \\
  $2$ & $5$ & $05.40$ & $5$ & $06.30$ \\
  $3$ & $5$ & $06.50$ & $5$ & $06.70$ \\
  $4$ & $5$ & $05.90$ & $5$ & $05.90$ \\
  $5$ & $5$ & $06.70$ & $5$ & $07.20$ \\
  $6$ & $5$ & $05.60$ & $5$ & $06.10$ \\
  $7$ & $5$ & $06.60$ & $5$ & $07.10$ \\
  $8$ & $5$ & $05.70$ & $5$ & $05.50$ \\
  $9$ & $5$ & $06.80$ & $5$ & $07.20$ \\
  $10$ & $5$ & $06.30$ & $5$ & $06.10$ \\
  \hline
  (c)
 \end{tabular}
 \begin{tabular}{ccccc}
  \hline
  Data & $\widehat{K}_n^s$ & $\%$ MA & $\widehat{K}_n$ & $\%$ MA \\
  \hline\hline
  $1$ & $5$ & $14.80$ & $2$ &  \\
  $2$ & $5$ & $14.20$ & $1$ &  \\
  $3$ & $5$ & $13.40$ & $2$ &  \\
  $4$ & $5$ & $13.60$ & $2$ &  \\
  $5$ & $4$ &  & $2$ &  \\
  $6$ & $5$ & $14.30$ & $1$ &  \\
  $7$ & $5$ & $15.10$ & $2$ &  \\
  $8$ & $5$ & $13.90$ & $2$ &  \\
  $9$ & $5$ & $14.70$ & $2$ &  \\
  $10$ & $5$ & $15.20$ & $1$ & \\
  \hline
  (d)
 \end{tabular}
 \\
 \begin{tabular}{ccccc}
  \hline
  Data & $\widehat{K}_n^s$ & $\%$ MA & $\widehat{K}_n$ & $\%$ MA \\
  \hline\hline
  $1$ & $5$ & $10.60$ & $3$ &  \\
  $2$ & $5$ & $11.30$ & $4$ &  \\
  $3$ & $5$ & $09.70$ & $3$ &  \\
  $4$ & $5$ & $09.60$ & $4$ &  \\
  $5$ & $5$ & $11.00$ & $4$ &  \\
  $6$ & $5$ & $10.50$ & $4$ &  \\
  $7$ & $5$ & $09.80$ & $4$ &  \\
  $8$ & $5$ & $10.70$ & $4$ &  \\
  $9$ & $5$ & $11.50$ & $4$ &  \\
  $10$ & $5$ & $12.50$ & $3$ & \\
  \hline
  (e)
 \end{tabular}
 \caption{\label{Tab3}{\small{For all these simulations, $L=10$ and $K=5$. In the tables (a), (b) and (c), we assumed $|S|=8$ and the $F_{ST}$ for the loci in $S$ were $0.0304$, $0.0355$ and $0.0407$ respectivelly. As expected, the increase of $F_{ST}$ for the loci in $S$ improves the performances of MixMoGenD. In the Tables (d) and (e), we assumed $|S|=6$, and the difference between running MixMoGenD with or without selection is clear. MA = Misassigned, $\widehat{K}_n^s$ and $\widehat{K}_n$ are the estimates of the number of populations with and without loci selection resprctively.}}}
\end{table}
\begin{table}[]
 \centering
 \begin{tabular}{c|ccccc}
  \hline
   & Pop1 & Pop2 & Pop3 & Pop4 & Pop5 \\
  \hline\hline
  Pop1 & $0.00000000$ & $0.04112990$ & $0.03024947$ & $0.02425668$ & $0.03535726$ \\
  Pop2 & $0.04112990$ & $0.00000000$ & $0.03831558$ & $0.02255300$ & $0.02756619$ \\
  Pop3 & $0.03024947$ & $0.03831558$ & $0.00000000$ & $0.02255183$ & $0.03251246$ \\
  Pop4 & $0.02425668$ & $0.02255300$ & $0.02255183$ & $0.00000000$ & $0.02509488$ \\
  Pop5 & $0.03535726$ & $0.02756619$ & $0.03251246$ & $0.02509488$ & $0.00000000$ \\
  \hline
 \end{tabular}
  \caption{\label{Tab4}{\small{The $F_{ST}$ between population $4$ and the others are all $<0.026$. MixMoGenD on this data set produces $4$ clusters and we observed that Pop4 was uniformly distributed in the $4$ clusters.}}}
\end{table}

\end{document}